\documentclass[12pt]{iopart}

\usepackage{graphicx}

\begin{document}
\title{Out-of-equilibrium Correlated Systems~: Bipartite Entanglement as a Probe of Thermalization}
\author{Didier Poilblanc}
\address{
Laboratoire de Physique Th\'eorique UMR5152, CNRS and Universit\'e de Toulouse, F-31062 France 
} 
\date{\today}
\begin{abstract}
Thermalization play a central role in out-of-equilibrium physics of ultracold atoms
or electronic transport phenomena. 
On the other hand, entanglement concepts have proven to be extremely useful to investigate quantum phases of matter.
Here, it is argued that {\it bipartite} entanglement measures provide key information on out-of-equilibrium states
and might therefore offer stringent thermalization criteria.
This is illustrated by considering a global quench in an (extended) XXZ spin-1/2 chain across its (zero-temperature) quantum critical point. A non-local {\it bipartition} of the chain {\it preserving translation symmetry} is proposed. The time-evolution  after the quench of the {\it reduced} density matrix of the half-system is computed and
its associated (time-dependent) entanglement spectrum is analyzed. 
Generically, the corresponding entanglement entropy 
quickly reaches a "plateau" after a short transient regime. However, in the case of the integrable XXZ chain, the low-energy entanglement spectrum still reveals strong time-fluctuations. In addition, its infinite-time average shows strong deviations from the spectrum of a Boltzmann thermal density matrix.
In contrast, when the integrability of the model is broken (by small next-nearest neighbor couplings),
the entanglement spectra of the time-average and thermal density matrices become remarkably similar.
\end{abstract}
\maketitle


\section{Introduction} 

Rapid progress in the field of ultracold atoms~\cite{ColdAtoms_Book} offer brend new perspectives to realize controlled 
experimental setup to investigate out-of-equilibrium physics.
Real-time observation of quantum dynamics of isolated systems have become possible~\cite{ColdAtoms_Isolated_Syst}. 
In addition, ultracold 
atoms loaded on optical lattices~\cite{ColdAtoms_Lattice,ColdAtoms_Book} 
or laser-cooled Coulomb crystals of charged ions~\cite{ColdIons} offer very clean experimental implementation
of simple lattice many-body Hamiltonians and provide simulators for Condensed 
Matter.
The ability to dynamically change parameters~\cite{ColdAtoms_Lattice2} in these Hamiltonians 
on short time scales could be exploited to realize quantum information 
processing~\cite{ColdAtoms_Book} or cooling~\cite{ColdAtoms_Cooling} devices.

In electronic condensed matter systems, relaxation towards steady states play a central role in many transport phenomena,
like e.g. in electric transport resulting from the application of a sudden voltage bias 
at the edges of a quantum dot~\cite{QuantumDot_Transport} or of a Hubbard chain~\cite{HubbardChain_Transport}.
Spin chains also offers simple generic systems to investigate out-of-equilibrium physics as e.g. heat transport~\cite{SpinChain_Heat_Transport}. 
However, conceptually, thermalization~\cite{Rigol_Nature} of non-equilibrium {\it isolated} quantum many-body systems after e.g.
a sudden change of Hamiltonian parameters (quantum quench) is still poorly understood, despite recent work 
on correlated bosons~\cite{Quench_BoseHubbard,Quench_HCB} in one-dimension (1D). 
Generally, whether some local observables approach steady values
and whether their time average 
equal the corresponding thermal average are often used as criteria of thermalization.
However, the fact that thermalization occur for certain local observables (according to the above criteria) 
does not at all  guarantee that other observables will also meet the criteria.

Independently, quantum information concepts have been applied with great success to several domains of Condensed Matter~\cite{RevModPhys},
giving new type of physical insights on exotic quantum phases. 
Quantum entanglement of a  A/B bipartition of  a many-body (isolated) quantum system can be characterized by 
the {\it groundstate} (GS) reduced density matrix (RDM) $\rho_A$ obtained from tracing out the B part.
The corresponding entanglement Von Neumann (VN) entropy $S_{\rm VN}=-{\rm Tr}\{\rho_A \ln\rho_A\}$
offers an extraordinary tool~\cite{EntanglementEntropy}, e.g. 
to identify underlying conformal field theory (CFT) structure in one-dimensional systems. 
Another central quantity is the {\it entanglement spectrum} (ES) defined by the (positive) eigenvalues of a 
(dimensionless) pseudo-Hamiltonian ${\cal H}$ defined by the implicit relation $\rho_A=\exp{(-{\cal H})}$. 
Remarkably, ES faithfully reflect CFT structures~\cite{OneDimension1},  topological symmetries~\cite{OneDimension2} or properties of {\it edge states} in fractional quantum Hall states~\cite{ES_FQHedges} or
low-dimensional quantum magnets~\cite{ES_ladder}. 

So far, time evolution of entanglement has been investigated only in very simple cases e.g. for 
a small segment in a 1D system after global or local quenches~\cite{Genway,RevModPhys,EntanglementEntropy}.
However, the full potential of entanglement measures has not been fully exploited to
investigate thermalization of many-body systems.   Because 
for a given {\it bipartition} (i.e. into two halves) of the whole system all the information of the GS is contained in its Schmitt decomposition 
and the ES is a (convenient) way of arranging the Schmitt coefficients, the ES contains the whole information of the state. 
The main goal here is therefore to use the bipartite ES to take the place
of local observables to investigate thermalization: 
whether the ES approach the spectrum of a thermal density matrix (whose effective temperature is fixed
by the excess energy brought in the quench) is therefore proposed as a stringent test of thermalization.
If the ES satisfies this thermalization
criterium, other observables would also do. 
The proposed choice of a bi-partition  of the total system is crucial since (i) an extensive subsystem is considered so that thermalization of local and {\it non-local}
observables is addressed at once, (ii) finite size scaling can be performed and (iii)
reduced density matrices benefit from numerous conserved quantities (total momenta and total spin) allowing for a direct comparison of the ES {\it separately in each symmetry sector} and hence an ultimate comparison between 
time-averaged and Boltzman thermal density matrices.
Note that non-local real-space partitions have also been used e.g. to define non-local order 
in {\it gapless} spin chains~\cite{NonLocalOrder}.  

Practically,  
a genuine correlated anisotropic 1D spin chain and the time evolution of its reduced 
density matrix, entanglement entropy and ES after a global quench  are considered using exact (Lanczos and full) diagonalization techniques. 
For the integrable case we have considered, the reduced density matrix exhibits strong time-fluctuations and its
infinite-time average significantly deviates from a thermal density matrix, despite the fact that the entanglement entropy reaches a well-defined entropy plateau. In contrast, thermalization, as defined by the above criteria, seems to 
be possible when integrability is broken by adding (small) extra terms to the Hamiltonian. 

\section{Model and setup}  

\begin{figure}
  \includegraphics[width=\columnwidth]{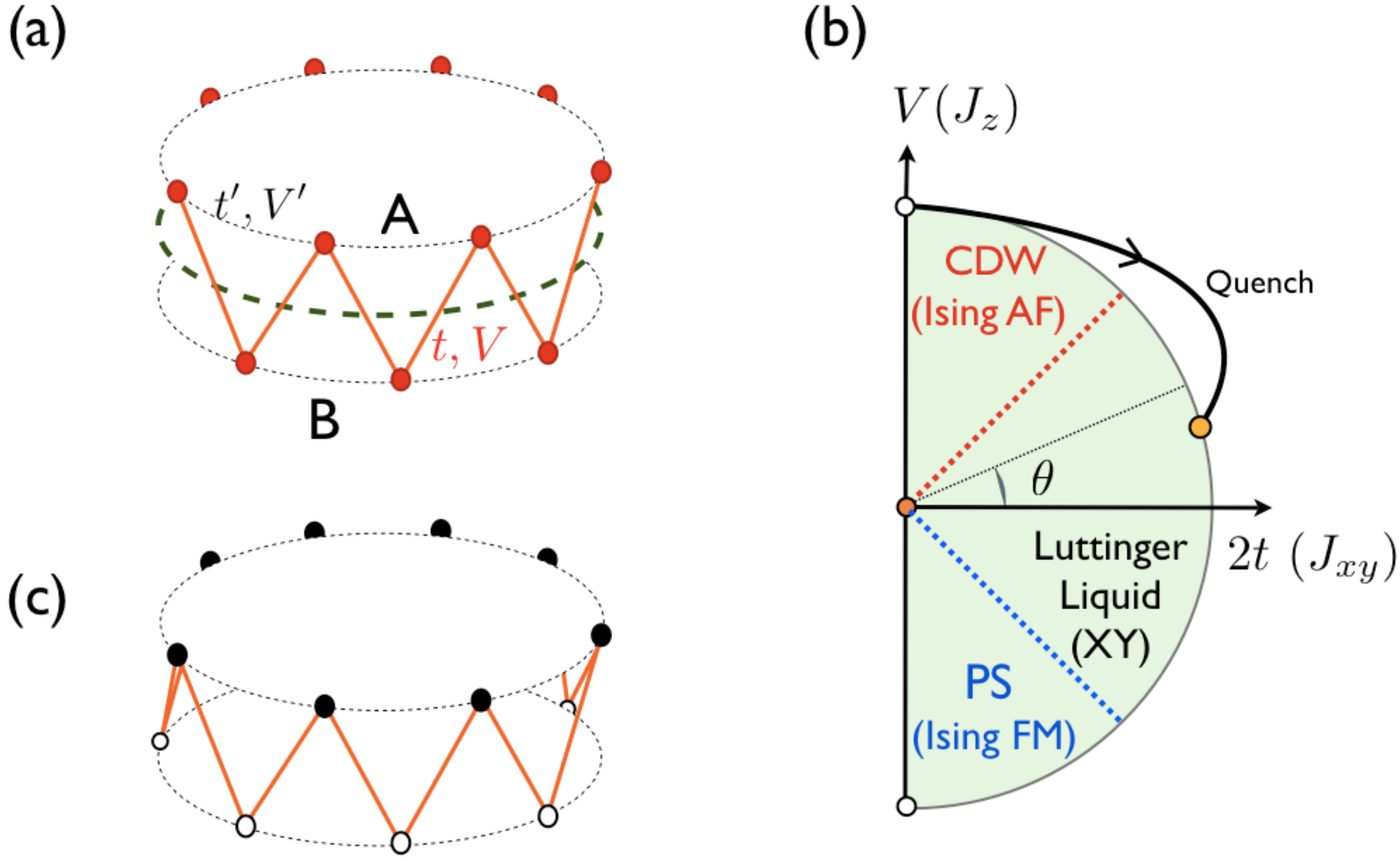}
  \caption{(Color online)
(a) A N-site (extended) XXZ spin-1/2 chain (drawn here as a "zig-zag" on a periodic ribbon) is partitioned 
into two identical A and B subsystems of $L=N/2$ sites by cutting bonds along the dashed line.
(b) Phase diagram of the XXZ spin (or hardcore boson) chain mapped onto a circle assuming $J_{xy}=\cos\theta$ and $J_{z}=\sin\theta$.
(c) The CDW groundstate for $\theta=\pi/2$ i.e. $t=0$ ($J_{xy}=0$) prepared before the quench~: all bosons (up spins) are located on A and the B sites
are empty (down spins). }
\label{Fig:xxz-phasediag}
\end{figure}

Let us now consider the 1D anisotropic spin-1/2 Heisenberg 
(so-called XXZ) model (Fig.~\ref{Fig:xxz-phasediag}),
\begin{equation}
H=J_z \sum_{i} S_i^z S_{i+1}^z + \frac{1}{2} J_{xy} (S_i^+ S_{i+1}^- + S_i^- S_{i+1}^+) \, ,
\end{equation}
whose (1D) parameter space can be mapped on a (half) unit circle 
assuming $J_{xy}=\cos\theta$ and $J_{z}=\sin\theta$.
Alternatively the system can be viewed as a (half-filled) hard-core boson chain with hopping 
$t=J_{xy}/2$ and nearest neighbor (NN) repulsion $V=J_z$. Its remarkable phase diagram 
obtained by Haldane~\cite{XXZ_PhaseDiagram} and shown in Fig.~\ref{Fig:xxz-phasediag}(b)
exhibits a Quantum Critical Point (QCP) located 
exactly at the SU(2)-symmetric point $\theta=\pi/4$ separating a gapped Charge Density Wave (CDW) 
insulating phase (Ising phase in the spin language) and a  gapless metallic Luttinger Liquid (XY phase). 
Phase Separation (PS) occurring for an attractive interaction $|V|>2t$ ($\theta<-\pi/4$) will not be of interest here. 
Ultimately, we shall consider adding next NN hopping $t'$ and repulsion $V'$ (see 
Fig.~\ref{Fig:xxz-phasediag}(a)) in order to break integrability.

\begin{figure}\begin{center}
  \includegraphics[width=0.8\columnwidth]{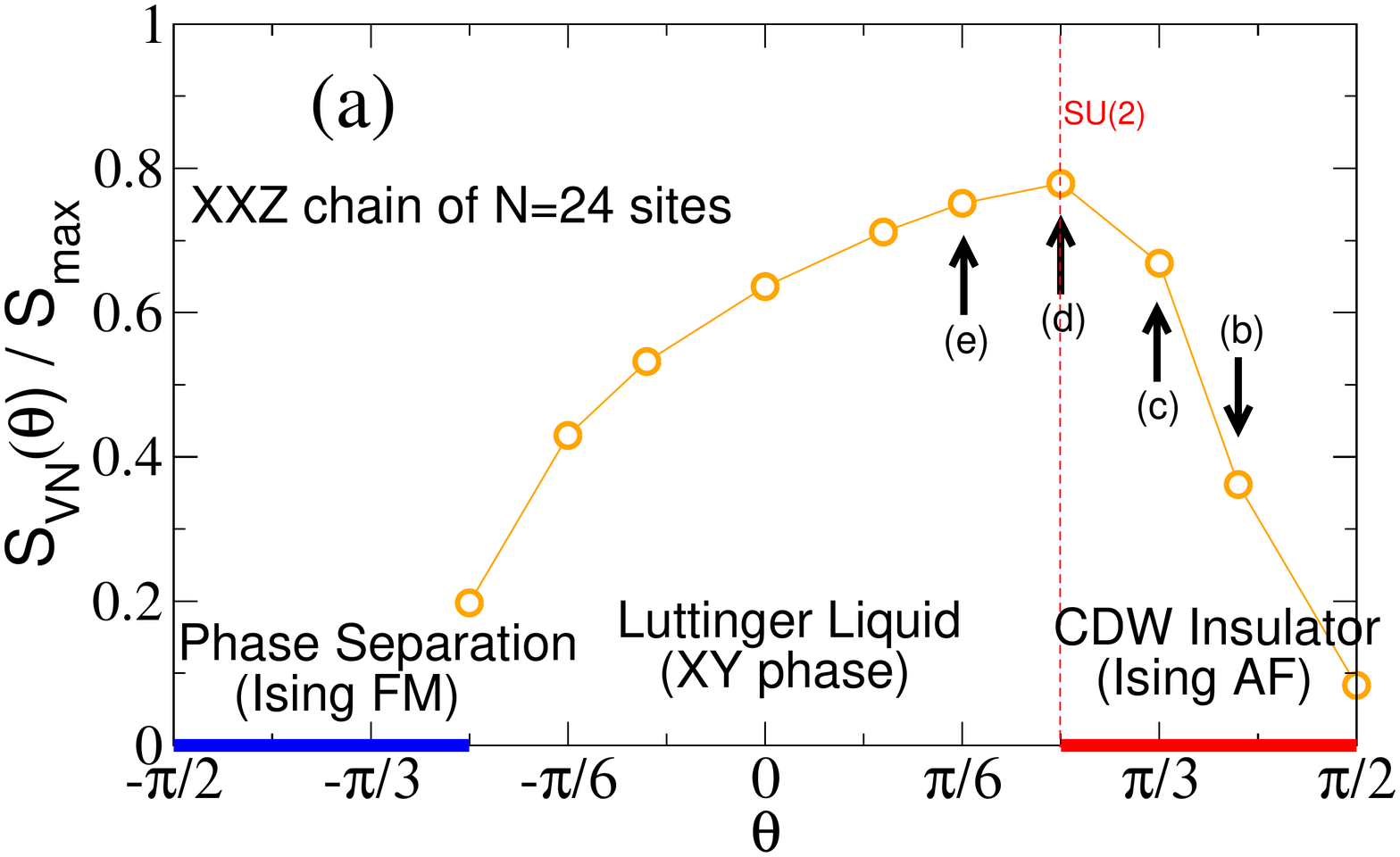}
 \begin{center}
 \vskip -1.8cm
  \includegraphics[width=\columnwidth]{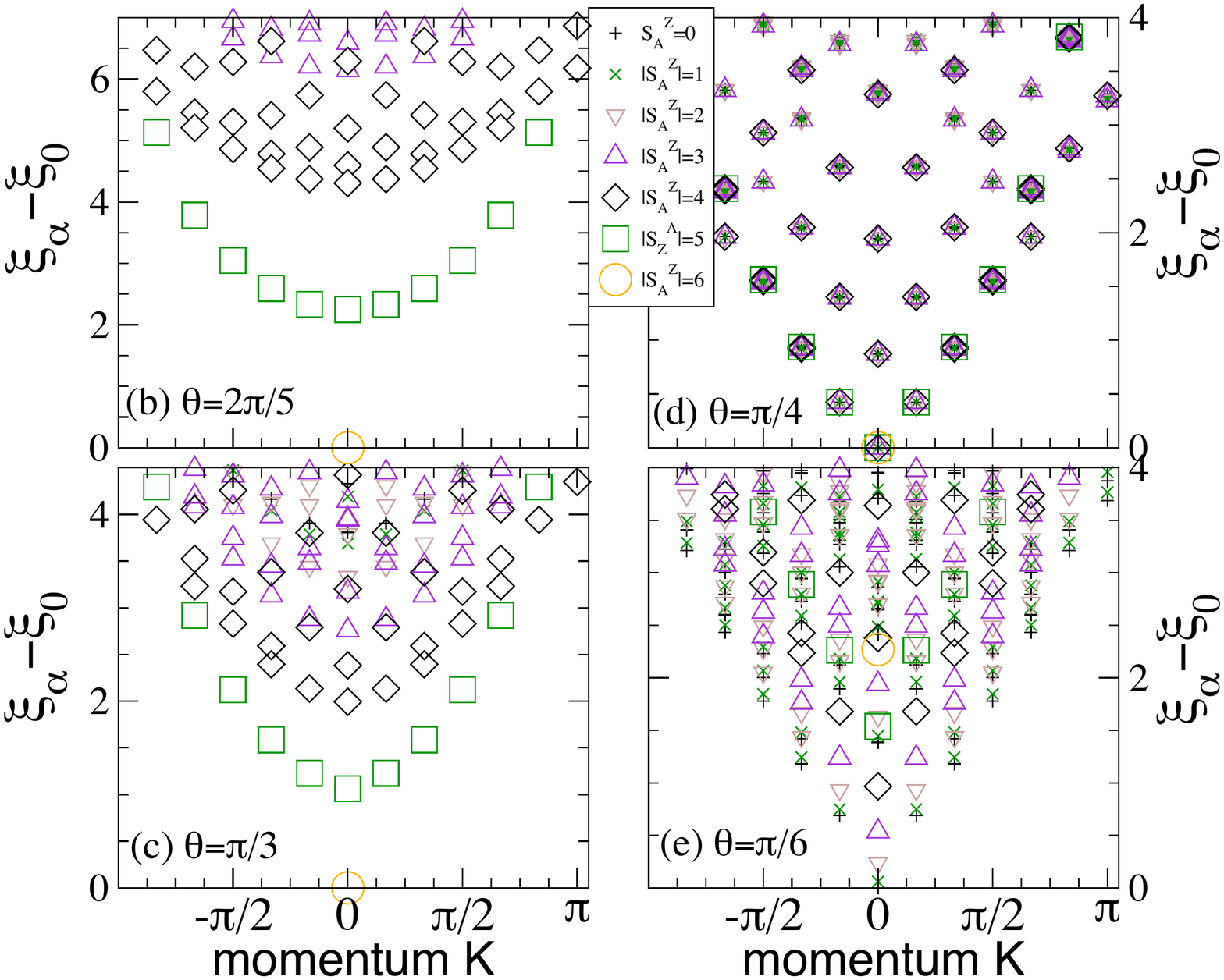}\end{center}  
  \caption{(Color online)
GS properties of the XXZ chain; (a) VN entanglement entropy vs $\theta$ computed on a N=24 site ring. The entropy is normalized by the maximum value $S_{\rm max}=L \ln{2}$
($L=N/2$).
(b-e) Typical entanglement {\it excitation} spectra (for the 4 values of $\theta$ shown in (b)) as a function of momentum $K$ along the ribbon. 
The eigenvalues $\xi_\alpha$ of ${\cal H}=-\ln{\rho_A}$ are labelled according to the Z-component of the total spin (i.e. the number of bosons $N_A=L/2-S_A^Z$) of the A  subsystem (legend of symbols on graph). Note the 
CDW  (LL ),  doubly-degenerate (unique) GS of ${\cal H}$ (of energy $\xi_0$)
carries $N_A=0$ or $N_A=L$ ($N_A=L/2$) hardcore bosons. }
\label{Fig:xxz-entanglement}\end{center}  
\end{figure}

The key to construct extensive quantities is to realize a non-local 
partition of the chain into "even" and "odd" sites. In other words, if the chain is drawn in a zig-zag fashion
as in Fig.~\ref{Fig:xxz-phasediag}(a), the A and B parts form the two edges of the 
system which  become explicit for $t'\ne 0$ and $V'\ne 0$.
One consider here finite chains of N ($=$ 16, 20 and 24) sites with periodic boundary conditions,
as shown in Fig.~\ref{Fig:xxz-entanglement}. The {\it groundstate} RDM of the 
subsystems $\rho_A=\rho_B$ can be computed using translation symmetry of each $L=N/2$ site
subsystem~\cite{ES_ladder} (each symmetry class is labelled by a momentum $K=2\pi n/L$) and 
the conservation of the Z-component of the total spin (i.e. the number of bosons), $S_A^Z+S_B^Z=0$.
Interestingly, liquid and insulating bosonic phases are characterized by qualitatively different entanglement properties. First, the entanglement entropy
{\it per unit length},
shown in Fig.~\ref{Fig:xxz-entanglement}(a) for a $N=24$ site chain, shows a (kink-like) maximum at the SU(2)-symmetric QCP and vanishes (in the thermodynamic limit) for the classical CDW (Ising) 
configuration obtained when $\theta\rightarrow \pi/2$.
Indeed, as shown in Fig.~\ref{Fig:xxz-phasediag}(c), for $\theta=\pi/2$ the (two-fold degenerate)
GS is a simple product of a completely filled (A or B) chain by a completely empty (B or A) one. 
Note that the symmetrized GS (i.e. the equal weight superposition of the two classical GS) 
still retains a $1/L$ finite size entropy as shown in Fig.~\ref{Fig:xxz-entanglement}(a). 
Secondly, each quantum phase is uniquely characterized by its ES defined as the spectrum
of ${\cal H}=-\ln{\rho_A}$: Fig.~\ref{Fig:xxz-entanglement}(b-c) 
(Fig.~\ref{Fig:xxz-entanglement}(d-e) ) for parameters in the CDW phase (LL phase) shows very distinctive 
features and a clear gapped (linear gapless) spectrum. In addition, 
at the QCP (Fig.~\ref{Fig:xxz-entanglement}(d))
one observes a SU(2) multiplet structure. 
 
\section{Time evolution and bipartite entanglement entropy}

Let us now consider a {\it sudden} quantum quench of the system at time $\tau=0$ (quasi-adiabatic quenches will be treated later on). For simplicity, the initial state $|\phi(0)\big>$ 
is chosen either (i) as one of the two (zero entropy) degenerate GS at $\theta=\theta_{\rm init}=\pi/2$ (i.e. for vanishing hopping) shown 
in Fig.~\ref{Fig:xxz-phasediag}(c),
where all hardcore bosons are located on a single edge, A or B, or (ii) as a symmetric combination of the two. At times $\tau>0$, the
boson hopping $t$ (spin-flip term in spin language) is switched on, i.e. the value of $\theta$ discontinuously jumps at $\tau=0$ to 
its "final" value $\theta_{\rm f}$ (see Fig.~\ref{Fig:xxz-phasediag}(b)). 

\begin{figure}\begin{center}
  \includegraphics[width=\columnwidth]{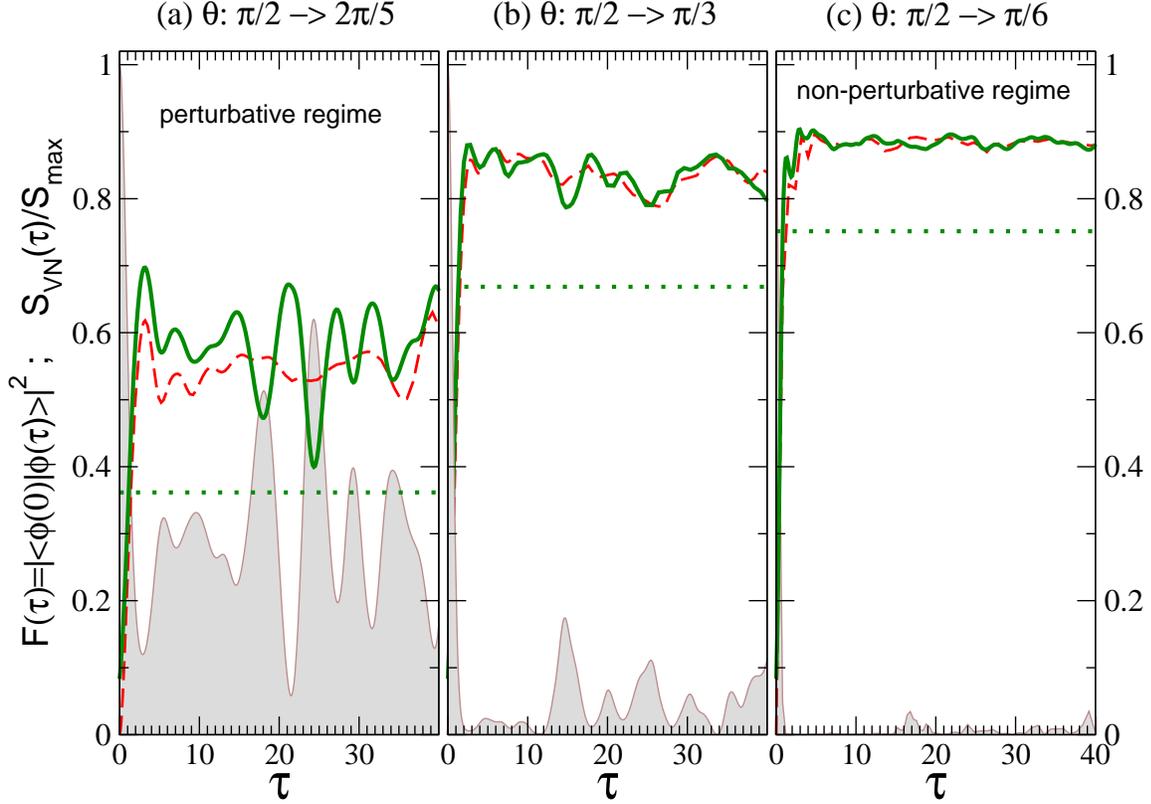}
    \caption{(Color online)
(a-c) Squared-fidelity (shaded) and VN entanglement entropy of various non-equilibrium states obtained after different {\it sudden} quenches  of the $N=24$ sites chain Hamiltonian,  
$\theta_{\rm init}\rightarrow\theta_{\rm f}$
as shown on plots ($t'=V'=0$). The continuous lines (dashed red line) correspond to a symmetric (non-symmetric) initial state (see text). The dotted (green) line is the 
GS entropy for $\theta=\theta_{\rm f}$.
}
\label{Fig:Entropy_TD}\end{center}  
\end{figure}

The time evolution of the system 
wavefunction, 
\begin{equation}
|\phi(\tau)\big>=\exp{(-i\tau H(\theta_{\rm f}))} |\phi(0)\big>
\end{equation}
 is easily computed by (arbitrary) time steps  of $\delta\tau$ (from $0.01$ to $0.8$) with arbitrary good precision~\cite{RouxThesis} (typically
better than $10^{-16}$) by Taylor expanding the time evolution operator $\exp{(-i\delta\tau H(\theta_{\rm f}))}$. Hence, for finite size systems under consideration, 
an {\it exact} computation of the time-dependent RDM, 
\begin{equation}
\rho_A(\tau)={\rm Tr}_B \,|\phi(\tau)\big>\big<\phi(\tau)|
\end{equation}
 can be done.
Results for the squared-fidelity $F(\tau)=|\big<\phi(0)|\phi(\tau)\big>|^2$ and the entanglement entropy, 
\begin{equation}
S_{\rm VN}(\tau)=-{\rm Tr}\{\rho_A(\tau) \ln\rho_A(\tau)\}
\end{equation}
are shown in Fig.~\ref{Fig:Entropy_TD} for increasingly "strong" quenches corresponding to $\theta_{\rm f}=2\pi/5$, $\pi/3$ and $\pi/6$. 
After a very short transient regime the squared-fidelity drops sharply and the entanglement entropy raises to a more or less well defined plateau, 
whose average value exceeds the value of the GS of the final Hamiltonian (f-GS). In the $N\rightarrow\infty$ limit, one expects a non-perturbative 
regime~\cite{NonPerturbative} where $F(\tau)\sim 2^{-N}\rightarrow 0$. However, on finite system and for small quench, $F(\tau)$
can remain large, as seen e.g. in Fig.~\ref{Fig:Entropy_TD}(a). In that regime, integrability of the model can 
play an important role~\cite{Dynamics_Integrable}. Therefore, from now on, one will assume a sufficiently 
large quench, let say $\theta_{\rm f}=\pi/6$, to observe time evolutions on finite clusters {\it generic of the thermodynamic limit}. 
This corresponds in fact to a quench "across" the QCP at $\theta=\pi/4$ into the region of stability of the LL phase. However, from 
Fig.~\ref{Fig:Entropy_TD}(c) we can see a well defined entropy plateau, suggesting that the system does reach a steady state after the quench.
Note also that one finds that non-symmetric and symmetrized initial states give very similar results
so that, for simplicity, one can restrict to the symmetric case hereafter.

The time evolution of any (static) observable $O$ of the A subsystem is given by $\tilde O(\tau)={\rm Tr}(\rho_A(\tau)O)$.
Time-average like ${\bar O}=\big<\tilde O\big>$, where 
$\big< G\big>\equiv\lim_{T\rightarrow\infty}\frac{1}{T} \int_{0^+}^T G(\tau) d\tau$,
can then be rewritten as ${\rm Tr}(\rho_A^{\rm ave}O)$, involving the {\it time-averaged} RDM
$\rho_A^{\rm ave}=\big<\rho_A\big>$.
Next, time-averages will be performed in a $\Delta\tau=40$ time interval using a mesh of 50 points (excluding the initial short transient regime).

\begin{figure}\begin{center}
  \includegraphics[width=\columnwidth]{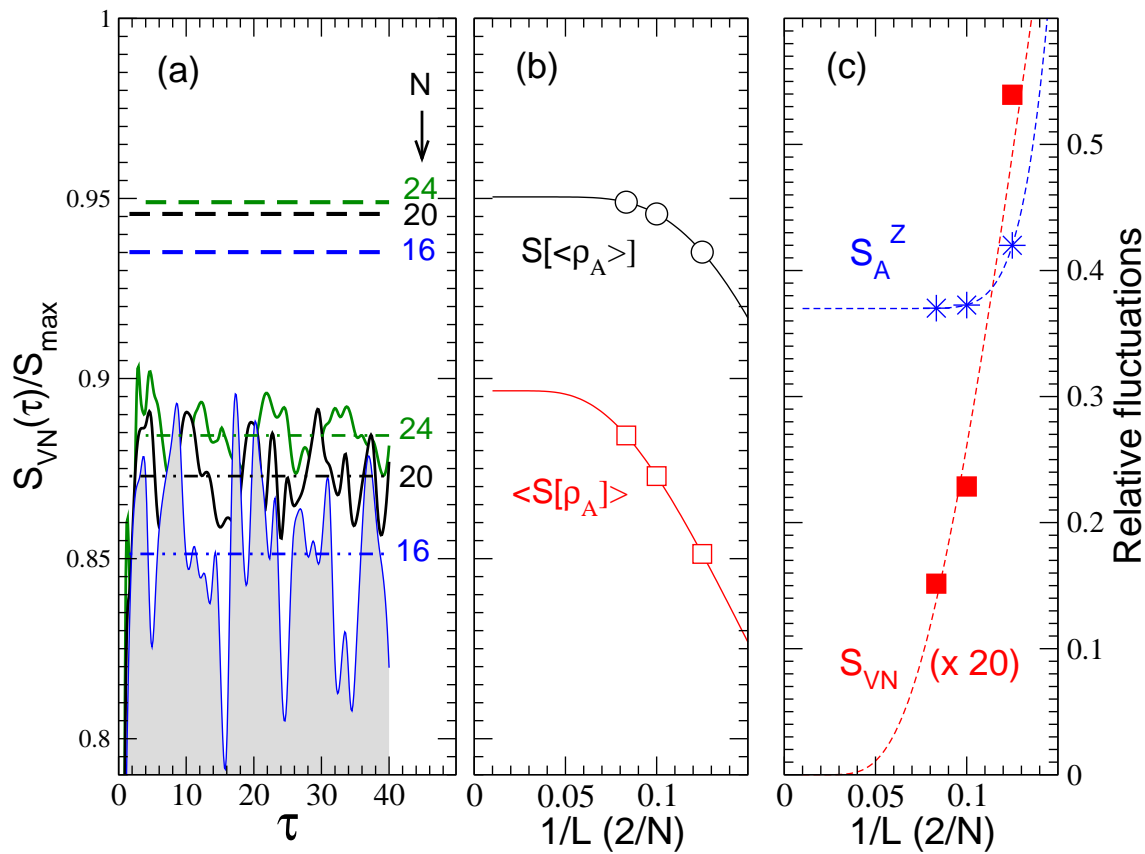}
  \caption{(Color online)
(a)  Time-dependent VN entanglement entropy after a {\it sudden} $\theta_{\rm init}=\pi/2\rightarrow\theta_{\rm f}=\pi/6$ quench ($t'=V'=0$), for different 
chain lengths $N$ as shown on plots. 
Dashed-dotted (dashed) lines corresponds to 
$\big< S_{\rm VN}\big>$ ($S_{\rm VN}^{\rm ave}$).
(b) Finite-size scaling of the averages shown in (a). (c) Finite-size scaling of the relative time fluctuations of the entanglement entropy 
(red squares) and quantum fluctuations of $S_A^Z$ (normalized by $L/2$).
Time-averages are all performed in a $\Delta\tau=40$ time interval using a mesh of 50 points. 
The data of b) and c) are well fitted assuming $\sim 2^{-\alpha L}$ finite size corrections.
}
\label{Fig:Entropy_scaling}\end{center}  
\end{figure}

Bipartite entanglement can provide precise characterization of the system
after the quench,  in the thermodynamic limit,
e.g. showing  whether or not it reaches a (quasi-) steady state.
One finds that the time fluctuations of $S_{\rm VN}(\tau)$ shown in Fig.~\ref{Fig:Entropy_scaling}(a)
vanish in the thermodynamic limit as revealed by the finite size scaling  of Fig.~\ref{Fig:Entropy_scaling}(c).
Interestingly, as seen in Fig.~\ref{Fig:Entropy_scaling}(a), we notice that the VN entropy of the average density matrix 
$\rho_A^{\rm ave}$, 
$S_{\rm VN}^{\rm ave}=-{\rm Tr}\{\rho_A^{\rm ave} \ln\rho_A^{\rm ave}\}$, differs 
from  $\big<S_{\rm VN}\big>$, the time-average of the entanglement 
entropy. A proper finite size scaling of the two quantities shown in Fig.~\ref{Fig:Entropy_scaling}(b)
proves that this fundamental property holds in the thermodynamic limit. 

Incidentally, it is also important to distinguish between "quantum" fluctuations 
$\overline{(O-\bar O)^2}={\rm Tr}\{ \rho_A^{\rm ave}(O-\bar O)^2\}$ and
"time" fluctuations $\big< (\tilde O-\bar O)^2 \big>$. An extreme example 
is the case of $S_A^Z$ where $\tilde{S_A^Z}(\tau)=0$ at all times 
(from the time-conserved $A\leftrightarrow B$ symmetry)
while quantum fluctuations are finite as shown in Fig.~\ref{Fig:Entropy_scaling}(c).

\section{Entanglement spectra: time evolution within the "entropy plateau"}

\begin{figure}\begin{center}
  \includegraphics[width=\columnwidth]{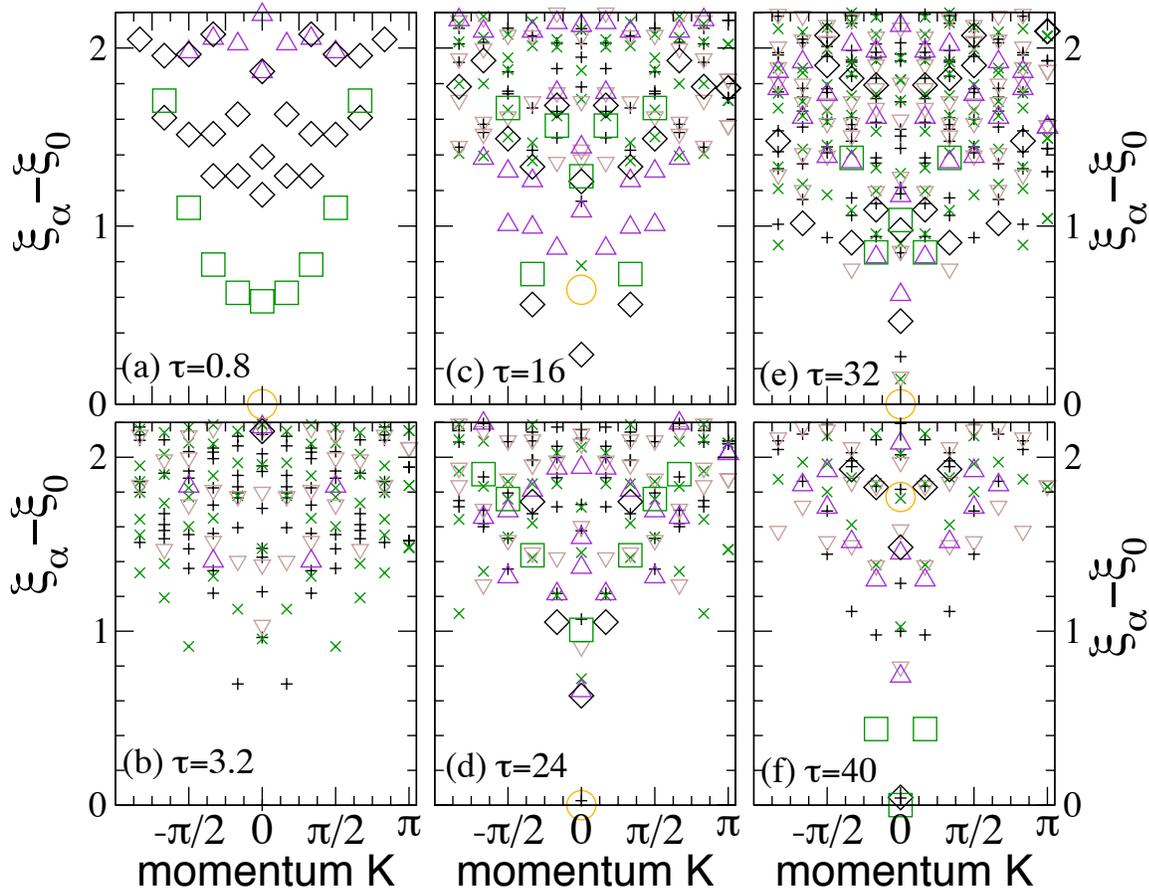}
  \caption{(Color online)
"Snapshots" at different times $\tau$ of the ES of  $\rho_A$. (a) lies within the transcient regime and (b-f) lie
within the "entropy plateau". 
 Symbols are similar to Fig.~\ref{Fig:xxz-entanglement}(b-e).
 }
\label{Fig:ES_vsK_vsTime}
  \end{center}  
\end{figure}

In order to understand the physical origin of the difference between $\big<S_{\rm VN}\big>$
and $S_{\rm VN}^{\rm ave}$ it is interesting to inspect more closely the behavior of $\rho_A(\tau)$ within the
"entropy plateau" regime as a function of time $\tau$. For this purpose, it is convenient to use
the (bipartite) {\it time-dependent} ES defined by the (positive) eigenvalues of the
(dimensionless) pseudo-Hamiltonian ${\cal H}(\tau)$ defined as,
\begin{equation}
\rho_A(\tau)=\exp{(-{\cal H} (\tau))} \, .
\end{equation}
As for the equilibrium GS, the spectrum of $\cal H$ is computed separately in every sector of the momentum 
$K=\frac{2\pi}{L/2}$ (using translation invariance of the A and B half-systems) and of the z-component of 
the total spin.
Fig.~\ref{Fig:ES_vsK_vsTime} shows "typical" entanglement spectra taken at different times. 
At short times, some memory of the initial state is clearly visible as in Fig.~\ref{Fig:ES_vsK_vsTime}(a).
At longer times, let's say $\tau>2.5$, it is remarkable to find very different spectra at different times although all such 
spectra lead roughly to a very comparable value of the entanglement entropy.  
In other word,  $\rho_A(\tau)$ fluctuates strongly in time on a {\it constant entropy} "hyper-surface" of the space of density matrices
with ${\rm Tr}\rho=1$.
This naturally explains why $\big<S_{\rm VN}\big>$
and $S_{\rm VN}^{\rm ave}$ differ substantially. Our finite size scaling suggests that this is a fundamental phenomenon and not a finite size effect. 

\section{Entanglement spectra: comparison between infinite-time average and thermal ensembles}

 The RDM $\rho_A$ contains all relevant information about the subsystem, much beyond any local observable. 
 As argued in the introduction, its associated ES 
 is therefore an ultimate observable to investigate thermalization. Furthermore, the existence of {\it conserved quantities} such as momentum and particle number 
makes the comparison between
time-average and thermal density matrices much finer, providing a stringent thermalization criterion (based on the
close similarity between these two quantities).

\begin{figure}\begin{center}
  \includegraphics[width=\columnwidth]{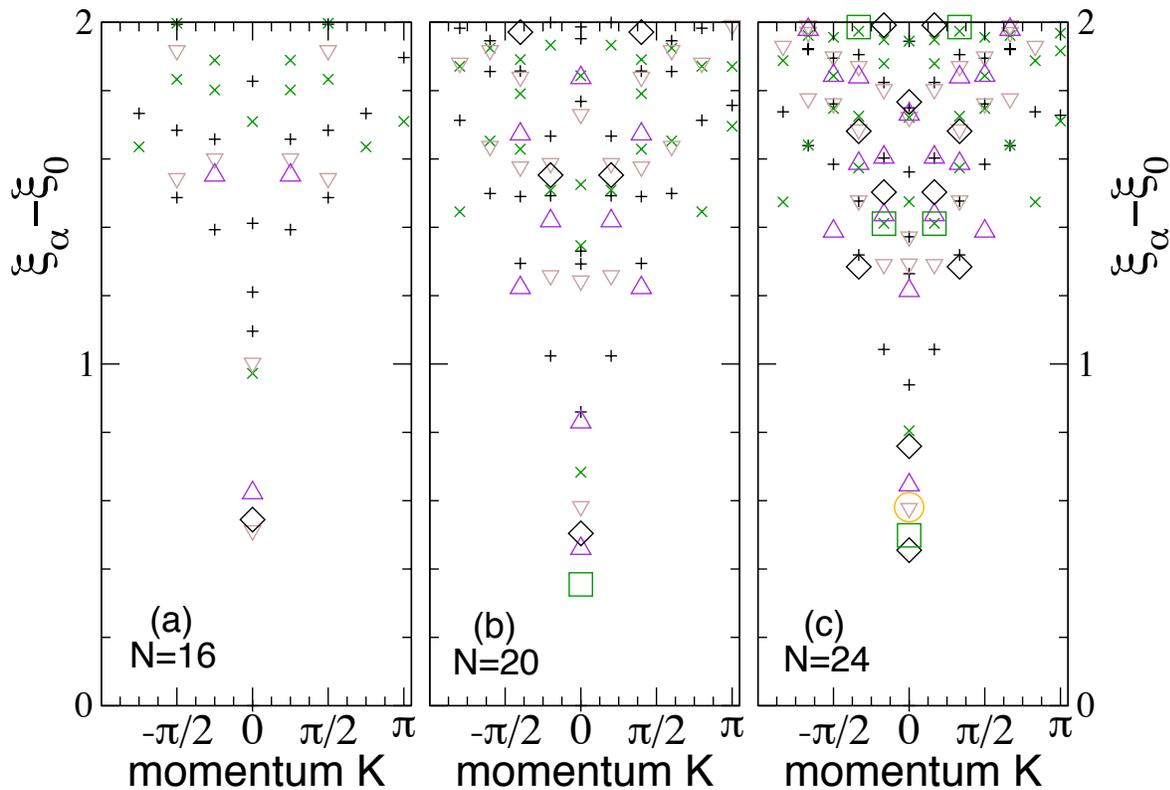}
  \caption{(Color online)
  Entanglement spectra of the {\it time-averaged} RDM obtained after a $\theta_{\rm init}=\pi/2\rightarrow\theta_{\rm f}=\pi/6$ quench
  in XXZ chains of different lengths, $N=16$, $20$ and $24$ ($t'=V'=0$).
 Symbols are similar to Fig.~\ref{Fig:xxz-entanglement}(b-e) and time-averages are performed as 
  for Fig.~\ref{Fig:Entropy_scaling}. 
}
\label{Fig:ES_scaling}\end{center}  
\end{figure}

The ES of $\rho_A^{\rm ave}$ computed for XXZ chains (in the entropy plateau) with $16$, $20$ and $24$ sites, and
shown in Fig.~\ref{Fig:ES_scaling}(a-c), reveal a smooth convergence with system size,
an accumulation of low pseudo-energy levels at $K=0$ and 
a small pseudo-energy gap, in sharp contrast with the ES of the f-GS shown in Fig.~\ref{Fig:xxz-entanglement}(e).  

\begin{figure}\begin{center}
  \includegraphics[width=\columnwidth]{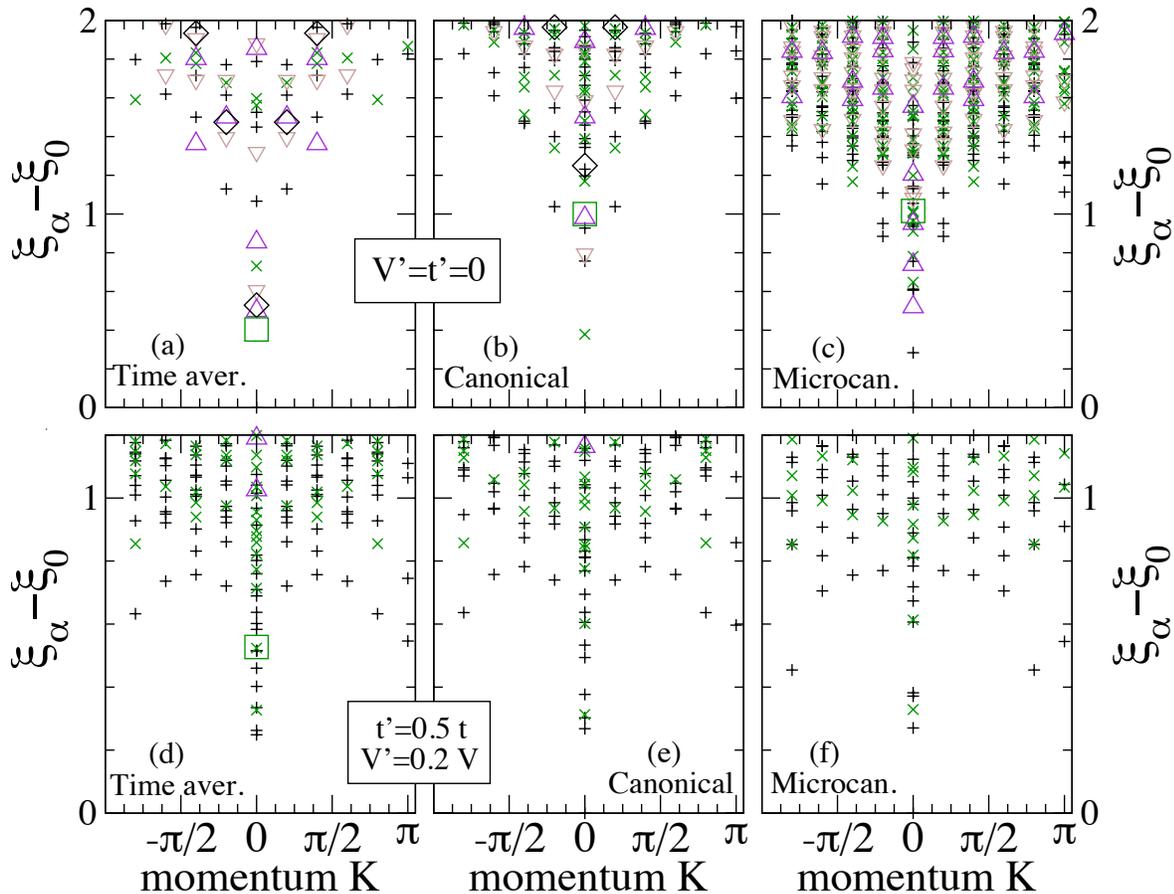}
  \caption{(Color online)
ES of  $\rho_A^{\rm ave}$ (a,d),  $\rho_A^{\rm can}$ (b,e) and  $\rho_A^{\rm micro}$ (c.f)
obtained by {\it full diagonalization} of $N=20$ site integrable ($t'=V'=0$) and non-integrable chains 
($\theta_{\rm init}=\pi/2\rightarrow\theta_{\rm f}=\pi/6$ quench).
 Symbols are similar to Fig.~\ref{Fig:xxz-entanglement}(b-e).
$1/\beta\simeq 2.07$ (b), $\Delta E=0.3$ (c), $1/\beta\simeq 1.48$ (e) and $\Delta E=0.1$ (f). }
\label{Fig:ES_Compare}\end{center}  
\end{figure}

In order to
probe thermalization,
we now wish to compare $\rho_A^{\rm ave}$ to the thermal average of $\rho_A$. Thermal averages can be defined as:
\begin{equation}
\rho_A^{\lambda}=\sum_\alpha^\prime w_\alpha^\lambda {\rm Tr}_B\, |\Psi_\alpha\big>\big<\Psi_\alpha|,
\label{thermal_density}
\end{equation}
where the prime means the sum is restricted to eigenstates of $H(\theta_f)$ 
with {\it non-zero} overlap with $|\phi(0)\big>$, accounting for conserved quantities like momentum and
z-component of total spin, and hidden conservation laws in the integrable case.
Both canonical ($w_\alpha^{\rm can}=\frac{1}{Z}\exp{(-\beta E_\alpha)}$ where $E_\alpha$ are
eigenenergies associated to $|\Psi_\alpha\big>$) and 
microcanonical ($w_\alpha^{\rm micro}={\rm Cst}$ in an energy window $[E_0-\Delta E,E_0+\Delta E]$) thermal ensembles
can be considered. Note, the effective temperature 
$1/\beta$ of the canonical ensemble is (implicitly) given by constraining the conserved mean-energy to be $E_0=-(V-V') L / 2$. Note that a full 
diagonalization of $H(\theta_f)$ is now required since the previous formula (\ref{thermal_density}) involves the complete
set of eigenstates $|\Psi_\alpha\big>$, hence limiting the available chain lengths to $N=20$.
The {\it infinite-time} average $\rho_A^{\rm ave}$ can be computed exactly also using (\ref{thermal_density}) 
with $w_\alpha^{\rm ave} =|\big< \Psi_\alpha |\phi(0)\big>|^2$ which, in contrast to the thermal averages, depends now
on the initial state. 
Incidentally, this enables to control the very good accuracy of the previous approximate averaging procedure,
as shown by a direct comparison between Fig.~\ref{Fig:ES_scaling}(b) and Fig.~\ref{Fig:ES_Compare}(a). 

The results shown in Figs.~\ref{Fig:ES_Compare} reveal striking differences between integrable and
non-integrable Hamiltonians.
For the latter,  surprisingly good agreement 
between the low-energy ES of $\rho_A^{\rm ave}$ and $\rho_A^{\rm can}$ is seen, suggesting that 
the {\it Eigenstate Thermalization Hypothesis}~\cite{Rigol_Nature} may apply up to the level of
the (bipartite) entanglement properties of eigenstates.
In contrast, for the integrable XXZ chain, noticeable differences between the various averages reveal incomplete 
thermalization which seems to be a fundamental property of such systems.

\section{Quasi-adiabatic quenches}

\begin{figure}\begin{center}
  \includegraphics[width=\columnwidth]{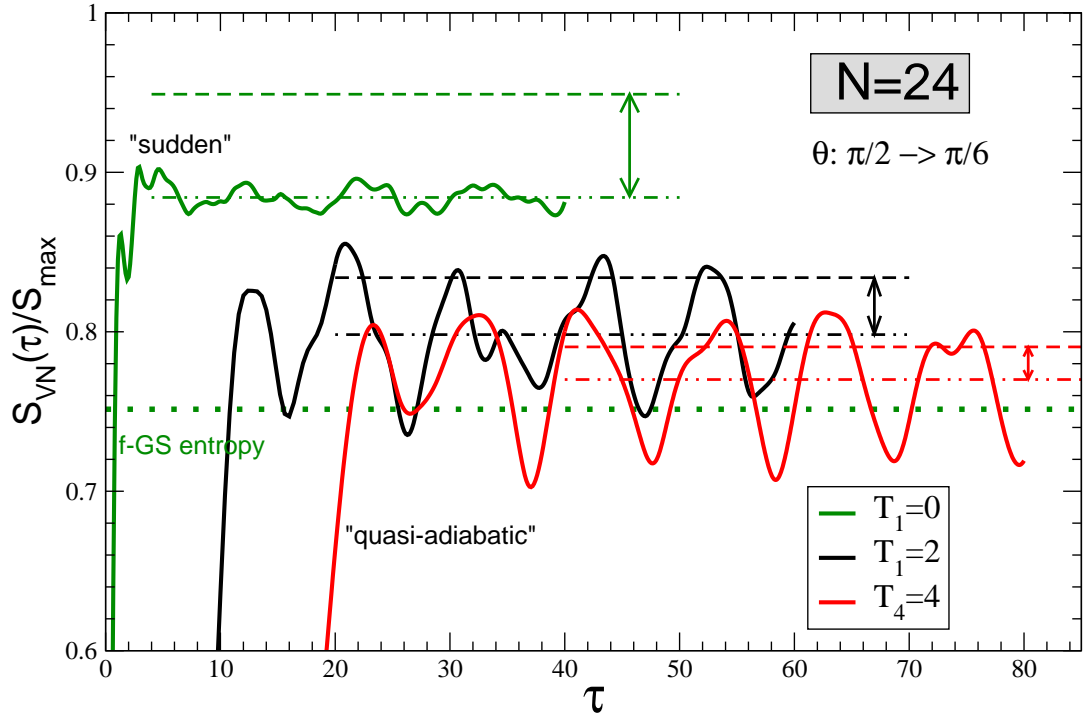}
  \caption{(Color online)
VN entanglement entropy of various non-equilibrium states obtained after  
a $\theta_{\rm init}=\pi/2\rightarrow\theta_{\rm f}=\pi/6$ quench realized on a 24-site XXZ chain ($t'=V'=0$)
over different time scales characterized by $T_1$. The arrows indicate the difference between the time-averaged entropy 
$\big< S_{\rm VN}\big>$ (dot-dashed lines) and the VN entropy of $\rho_A^{\rm ave}$, $S_{\rm VN}^{\rm ave}$ (dashed lines). Time-averages have been performed using 50 bins in a $\Delta\tau=40$ time-interval
within the plateaux.
}
\label{Fig:Entropy_TD2}\end{center}  
\end{figure}

The system after the quench possesses an excess of entanglement entropy (per unit length) compared to the f-GS.  
There is in fact an interesting correlation between this excess entropy and the "speed" at which the quench is performed.
A smooth quench can be realized by considering a time-dependent Hamiltonian 
$H(\theta(\tau))$ where $\theta(\tau)$ decreases continuously from 
$\theta_{\rm init}=\pi/2$ to $\theta_{\rm f}=\pi/6$
in a time-interval $[0,T_{\rm f}]$. One can choose e.g. 
\begin{eqnarray}
\theta(\tau)=(1-\tilde w_\tau)\theta_{\rm init}   + \tilde w_\tau\theta_{\rm f} ,\\
 \tilde w_\tau=(w_\tau-w_0)/(w_{\rm T_f}-w_0),\\
 w_\tau=1/(1+\exp{((T_{\rm f}/2-\tau)/T_1)}),
\end{eqnarray}
and $T_1=T_{\rm f}/10$.
As shown in Fig.~\ref{Fig:Entropy_TD2}, under increasing the characteristic time $T_1$ the average value of the entropy plateau decreases towards the GS value, as expected for increasingly adiabatic processes.
Interestingly, the deviation $\big< S_{\rm VN}\big> -S_{\rm VN}^{\rm ave}$ also simultaneously decreases (to reach zero for a fully adiabatic process). 

\begin{figure}\begin{center}
  \includegraphics[width=\columnwidth]{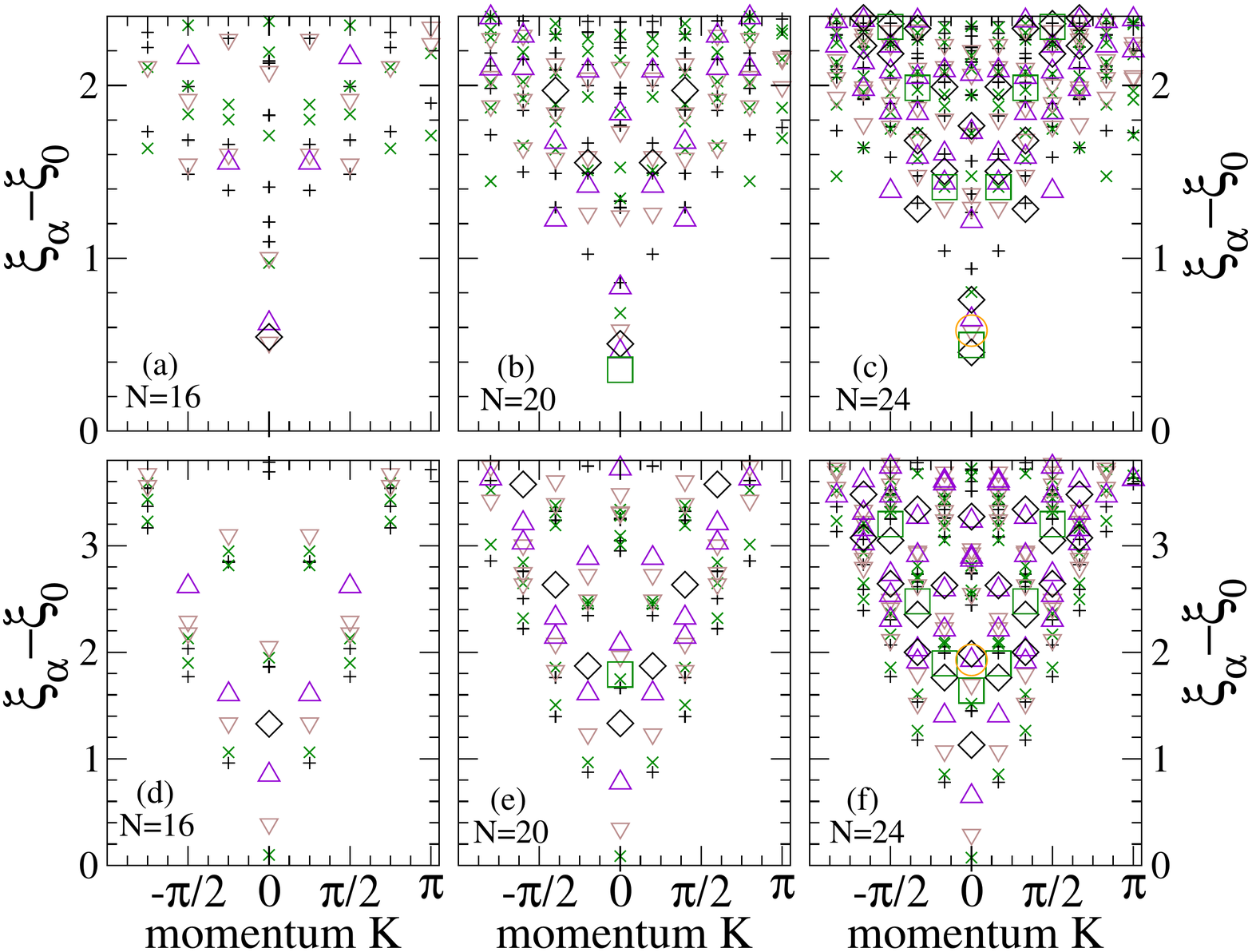}\end{center}  
  \caption{(Color online)
  Entanglement spectra of {\it time-averaged} reduced density matrices
  in XXZ chains of different lengths, $N=16$, $20$ and $24$ ($t'=V'=0$). 
  Comparison between a sudden quench (a-c) (data similar to Fig.~\ref{Fig:ES_scaling} reproduced here for convenience) 
  and a quasi-adiabatic $T_1=4$ quench (d-f).
  In both cases, $\theta_{\rm init}=\pi/2\rightarrow\theta_{\rm f}=\pi/6$. 
  }
\label{Fig:ES_adiab}
\end{figure}

Sudden and continuous quenches also give very different ES of the time-averaged reduced density matrix
$\rho_A^{\rm ave}$. 
Data for continuous and sudden quenches are compared in Fig.~\ref{Fig:ES_adiab}. 
Remarkably, the spectra in the quasi-adiabatic case resemble very closely the equilibrium
f-GS spectrum of Fig.~2(e) of the main paper with a clear linear envelope, in contast to the spectra 
obtained for a sudden quench. However, we note that the requirement (e.g. on the time scale $T_1$) to have a
quasi-adiabatic process 
becomes more and more stringent for increasing system size because of the vanishing of the finite-size gap (in the 
XY phase).

\section{Summary}

To summarize, the concept of bipartite entanglement is introduced in a genuine out-of-equilibrium 
{\it isolated} many-body system. I propose that it provides a simple and complete probe of thermalization, 
beyond the investigation of simple local observables. 
In this framework, absence of thermalization is diagnosed whenever the reduced density matrix deviates (once averaged
over a sufficiently large time interval) noticeably from the thermal (reduced) density matrix taken at an effective temperature imposed
only by the initial conditions. 
Generically, we find that the system approaches quickly an entanglement entropy plateau.
However, integrable and non-integrable chains behave quite differently within this plateau. 
The RDM of the integrable chain strongly fluctuates in time and does not thermalize
according to the above criterium. 
This is to be contrasted to the case where extra terms are introduced in the chain Hamiltonian to break the 
integrability: in such a case, the ES of the RDM and the spectrum of the thermal density matrix become very similar,
a strong hint that thermalization can occur up to the level of an extensive subsystem even though the 
full system is completely isolated.

{\it Acknowledgements} -- I thank D.~Braun, B.~Georgeot, T.~Lahaye, I.~Nechita, P.~Pujol, N.~Regnault, M.~Rigol,
G.~Roux and K.~Ueda for interesting discussions,  
IDRIS (Orsay, France) for CPU time on the NEC-SX8 supercomputer and the French Research Council (ANR) for funding. 

 
\end{document}